\documentclass[12pt,preprint]{aastex}
\shorttitle{Bullet Cluster in cDE Models}
\shortauthors{Lee \& Baldi}
\begin{document}
\title{BREAKING THE COSMIC DEGENERACY BETWEEN MODIFIED GRAVITY AND MASSIVE NEUTRINOS WITH THE COSMIC WEB}
\author{Junsup Shim\altaffilmark{1}, Jounghun Lee\altaffilmark{1}, 
Marco Baldi\altaffilmark{2,3,4}}
\altaffiltext{1}{Astronomy Program, Department of Physics and Astronomy, FPRD, 
Seoul National University, Seoul 151-747, Korea \email{jsshim@astro.snu.ac.kr, jounghun@astro.snu.ac.kr}}
\altaffiltext{2}{Dipartimento di Fisica e Astronomia, Alma Mater Studiorum Universit\`a di Bologna, viale Berti Pichat, 
6/2, I-40127 Bologna, Italy}
\altaffiltext{3}{INAF - Osservatorio Astronomico di Bologna, via Ranzani 1, I-40127 Bologna, Italy}
\altaffiltext{4}{INFN - Sezione di Bologna, viale Berti Pichat 6/2, I-40127 Bologna, Italy}
\begin{abstract}
In a recent work, Baldi et al. highlighted the issue of cosmic degeneracies, consisting in the fact that
the standard statistics of the large-scale structure might not be sufficient to conclusively 
test cosmological models beyond $\Lambda $CDM when multiple extensions of the standard scenario coexist in 
nature. In particular, it was shown that the characteristic features of an $f(R)$ Modified Gravity theory and of
massive neutrinos with an appreciable total mass $\Sigma _{i}m_{\nu _{i}}$ are suppressed in most of the basic 
large-scale structure observables for a specific combination of the main parameters of the two non-standard models.
In the present work, we explore the possibility that the mean specific size of the supercluster spines 
-- which was recently proposed as a non-standard statistics by Shim and Lee to probe gravity at large scales -- can 
help to break this cosmic degeneracy. By analyzing the halo samples from N-body simulations featuring 
various combinations of $f(R)$ and  $\Sigma _{i}m_{\nu _{i}}$ we find that -- at the present epoch -- the value of 
$\Sigma _{i}m_{\nu _{i}}$ required to maximally suppress the effects of $f(R)$ gravity on the specific sizes of the 
superclusters spines is different from that found for the other standard statistics. Furthermore, it is also shown that at 
higher redshifts ($z\ge 0.3$) the deviations of the mean specific sizes of the supercluster spines for all of the four 
considered combinations from its value for the standard $\Lambda$CDM case are statistically significant. 
\end{abstract}
\section{INTRODUCTION}

The clustering of the large-scale structures (LSS, hereafter)  and its evolution with redshift encode information about
the statistical properties of the initial conditions of the universe and the laws governing the gravitational instability 
processes that determine the growth of primordial density perturbations.  
The basic LSS statistics, such as the two point-correlation function of the density field, the mass function of 
galaxy clusters and the halo bias, are frequently employed to quantify the main properties of the LSS clustering,
thereby providing a handle on the underlying cosmological model. Such basic statistics have played a vital role in 
establishing the concordance $\Lambda $CDM scenario by allowing to place tight constraints on its basic cosmological 
parameters.

Furthermore, the standard LSS statistics are routinely employed to test not only the key cosmological parameters of 
the $\Lambda$CDM cosmology \citep[e.g.,][and references therein]{addison-etal13,didio-etal14} 
but also the viability of alternative cosmological models, such as e.g. the Warm Dark Matter model 
\citep[WDM, see e.g.][]{SM11,viel-etal12}, the coupled Dark Energy scenario 
\citep[cDE, see e.g.][]{moresco-etal13}, and various Modified Gravity theories
\citep[MG, see e.g.,][]{SR08,SK09,stril-etal10,lombriser-etal12,abebe-etal13}. 

The main motivation behind most of these alternative models was to overcome the observational and  theoretical 
shortcomings of the $\Lambda$CDM cosmology. For instance, the free-streaming effect of the WDM particles  
has been invoked as a possible solution to the tension between the $\Lambda$CDM predictions and astrophysical 
observations on (sub-)galactic scales \citep[e.g.,][and references therein]{menci-etal12}, even though recent constraints from the 
Lyman-alpha forest seem to exclude the WDM particle mass range required to address the tension \citep[][]{viel-etal13}.  
Similarly, the dark sector interactions that characterise the cDE scenario can alleviate the fine-tuning problems
of the cosmological constant \citep[e.g.,][]{wetterich95,amendola00,amendola04}, while possible large-scale 
deviations of the laws of gravity from their standard GR form can accommodate the present acceleration of the 
universe without requiring dark energy with negative pressure 
\citep[for a comprehensive review, see][]{mg_review12}. 

However, some cautionary remarks have been recently raised on the rosy prospects for the basic statistics of LSS as 
a powerful discriminator of alternative cosmologies. \citet{wei-etal13} theoretically proved that the WDM, cDE 
and MG models are hard to be differentiated from one another just by tracing the expansion and growth history 
with LSS observations. Additionally,  it was shown that the presence of a cosmological background of massive 
neutrinos might significantly suppress the main observational footprints of various cDE and MG models 
\citep[e.g.,][]{lavacca-etal09,motohashi-etal13,he13}. 

Such early predictions based on linear observables have been recently 
confirmed and extended to the non-linear regime of structure formation by \citet{baldi-etal14}, who studied the 
simultaneous effect on structure formation of a $f(R)$ MG model and of massive neutrinos  
(the ``$f(R)+\nu$" model, hereafter) by means of large $N$-body simulations. 
The $f(R)$ gravity is one of the viable and most widely investigated MG theories 
\citep[see e.g.,][and references therein]{lombriser14} where the Ricci scalar $R$ is replaced by a function $f(R)$ 
in the Einstein-Hilbert action. With suitable choices of such function, the model can be tuned to match the same 
expansion history of the standard $\Lambda $CDM cosmology \citep[][]{HS07}. Nonetheless, the 
derivative of the $f$ function $df/dR$ -- which represents an additional degree of freedom called ``the scalaron" --  is 
expected to mediate a {\it fifth-force} \citep[see][for a review]{DT10,SF10}, implying that the evolution of density 
perturbations will be different as compared to $\Lambda $CDM even for an identical expansion history. The viability of 
the model is ensured by the Chameleon screening mechanism  \citep[e.g.][]{KW04,brax-etal08} that allows to recover 
the behaviour of standard GR in overdense regions of the Universe.

In their recent work, \citet{baldi-etal14} noted that for a given $f(R)$ model there is a specific value of the total neutrino 
mass $\Sigma_{i}m_{\nu _{i}}$ that cancels the effects of the Modified Gravity in most of the standard basic LSS 
observables, thereby yielding -- besides an identical expansion history -- also the same LSS statistics as 
$\Lambda$CDM at the present level of observational accuracy. In other words, the free-streaming effect of massive 
neutrinos \citep{LP06} effectively cancels out that of the fifth-force of the $f(R)$ gravity on the large scale structure. 
Calling it a ``cosmic degeneracy", \citet{baldi-etal14} regarded this result as an indication of the fundamental limitation 
of the basic statistics of LSS as a test of alternative cosmologies, concluding that some novel  independent statistics or 
some other independent constraints (as e.g. a laboratory determination of the neutrino mass) are 
necessary to break the degeneracy. 

Meanwhile, \citet{SL13} have recently developed a new diagnostic based on the filamentary cosmic web for testing 
gravity at large scales. Such new approach has shown that the filamentary pattern of the cosmic web is significantly 
affected by deviations from the standard gravitational behaviour, either in terms of MG or cDE models.  More 
specifically, considering the supercluster spines (i.e. the main stems of the superclusters) as the richest 
filamentary structures in the cosmic web, \citet{SL13} determined their specific sizes to quantify 
the degree of the straightness of the superclusters, finding that the specific size distributions of the supercluster 
spines substantially differ among various cDE models \citep[see again][]{SL13}, implying it being indeed a good 
indicator of cDE.  Similarly, in a subsequent paper \citet{shim-etal14} also investigated the effect of $f(R)$ gravity on 
the specific sizes of the supercluster spines and showed that the evolution trends of the specific size distributions of 
the superclusters differ between the cDE and the $f(R)$ gravity models, which indicated that this new diagnostic is in 
principle capable of  breaking the degeneracy between the two models. 

In the light of the works of \citet{shim-etal14} and \citet{baldi-etal14}, in the present study we aim 
to explore  whether the the degree of the straightness of the superclusters quantified by the specific size of the 
supercluster spines can be also useful for breaking the cosmic degeneracy between the $\Lambda$CDM and the 
$f(R)+\nu$ models. In section \ref{sec:data} we will describe how the supercluster samples are obtained from the 
simulation datasets obtained from the work of \citet{baldi-etal14}. In section \ref{sec:result} we will show how the 
specific sizes of the supercluster spines are affected by the simultaneous effects of the $f(R)$ gravity and the massive 
neutrinos. In section \ref{sec:con} we discuss the implications of our results as well as possible prospects for future further 
improvements. 
 
\section{CONSTRUCTING THE SUPERCLUSTER SAMPLES}\label{sec:data}

\citet{baldi-etal14} carried out large N-body simulations for the standard GR+$\Lambda$CDM  and also for 
a $f(R)$ MG model with four possible different values of the total neutrino mass $\Sigma _{i}m_{\nu _{i}}$, namely 
$\left\{ 0\,, 0.2\,, 0.4\,, 0.6\right\}$ eV. The specific $f(R)$ model considered in the simulations corresponds to the 
widely investigated parameterization by \citet{HS07}, for $n=1$ and for a scalar amplitude at the present epoch of 
$f_{R0}\equiv df/dR\vert_{t_{0}}=-10^{-4}$. Although recent observations indicate $\vert f_{R0}\vert\le 10^{-5}$ 
on the cluster scale \citep[e.g.,][]{schmidt-etal09,lombriser-etal12}, such a large value of 
$\vert f_{R0}\vert=10^{-4}$ was chosen on purpose in order to maximize the effects of MG and to emphasize how even 
for large deviations from standard GR the cosmic degeneracy between $f(R)$ and massive neutrinos can very 
effectively suppress the observational signatures of both non-standard models. Furthermore, the above-mentioned 
observational constraints on $f(R)$ gravity have been derived assuming massless neutrinos, and might therefore 
become looser (again as a result of the degeneracy between MG and massive neutrinos) 
if the total neutrino mass is allowed to vary as a free parameter.

The simulations have been carried out by means of a suitable combination of the {\small MG-GADGET} code by 
\citet{puchwein-etal13} (specifically designed for MG cosmologies) and of the particle-based implementation of 
massive neutrinos by \citet{viel-etal10}. Both algorithms are independent modules of the widely-used TreePM 
N-body code {\small GADGET} \citep{gadget2}. With such combined code at hand, \citet{baldi-etal14}
followed the evolution of a cosmological volume of $1 h^{-3}$ Gpc$^{3}$ filled with $2\times 512^{3}$ particles 
for the CDM and neutrino components, with standard cosmological parameters consistent with the latest results of the 
Planck satellite mission \citep[][]{planck-etal13}, from $z=99$ to $z=0$. A catalog of CDM particle groups has 
been compiled on-the-fly for several different redshifts by means of a standard Friends-of-Friends (FoF) algorithm with 
a linking length set to the conventional value of $20\%$ of the mean inter-particle separation. For a more detailed 
description of the simulations, we refer the interested reader to the \citet{baldi-etal14} paper. 

From the cluster-size halos selected from the FoF catalogs at three different epochs 
($z=0,\ 0.3,\ 0.6$), we extract the superclusters following the approach described in \citet{SL13} and 
\citet{shim-etal14}. First of all, a sample of the cluster-size halos is generated by choosing those 
FoF halos with masses above a lower threshold of $10^{13}\,h^{-1}M_{\odot}$, at each redshift and 
for each model. The mean separation distance among such cluster-size halos is then calculated and the marginally-bound 
superclusters are identified by running again a FoF group finder on the positions of the cluster-size halos with 
the linking length parameter of $30\%$ the mean inter-halo separation. Finally, by binning the halo mass range of the 
two catalogs and calculating their number densities within each mass bin, we determine both the cluster and the 
supercluster mass functions.

\begin{figure}[ht]
\begin{center}
\plotone{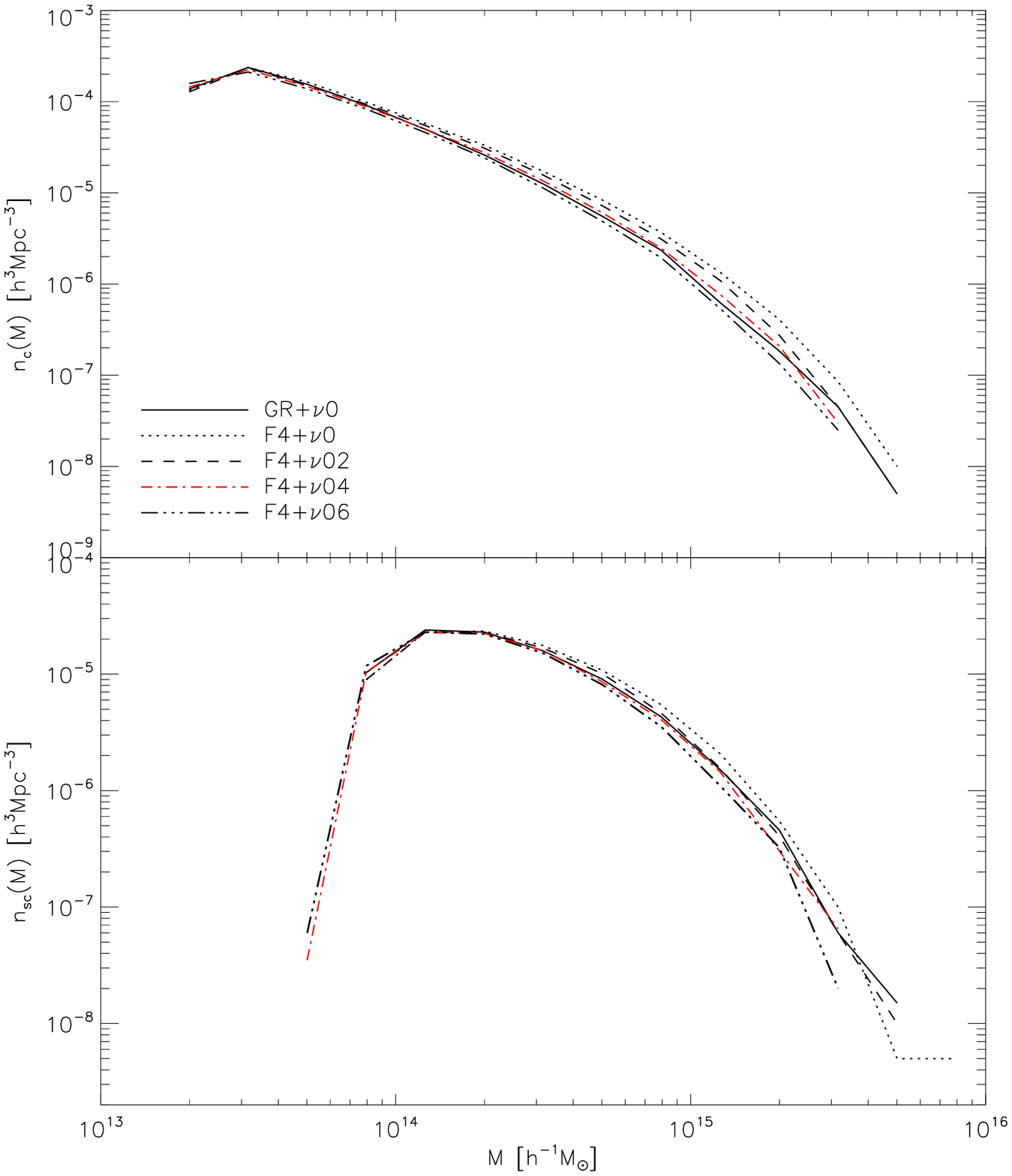}
\caption{Mass functions of the cluster halos and the superclusters for five different models at $z=0$ in the top and 
the bottom panel, respectively. In each panel the (red) dot-dashed line corresponds to the {\it most degenerate} 
combination of the models in the standard LSS statistics.}
\label{fig:mf}
\end{center}
\end{figure}

The top and the bottom panels of Figure \ref{fig:mf} show, respectively, the cluster and the supercluster mass 
functions at $z=0$ for all the five models under investigation.  
As can be clearly seen in the plots, a lower value of $\Sigma _{i}m_{\nu _{i}}$ generally exhibits a higher number 
density of high-mass superclusters  ($M\ge 10^{15}\,h^{-1}M_{\odot}$) . Note also that both the cluster and the supercluster mass 
functions for the $\Lambda$CDM model are almost identical to those for the F4+$\nu{04}$ model (red dot-dashed line), which 
indicates that the neutrino mass required to cancel out the effect of the fifth force on both observables is of about $0.4\,$eV. This result 
is in full agreement with the findings of \citet{baldi-etal14} that the standard LSS statistics including the cluster mass functions are not 
capable of discriminating the $\Lambda$CDM model  from the F4+$\nu{04}$ model.

\section{SIMULTANEOUS EFFECT OF MG AND $\nu$ ON THE SUPERCLUSTER STRAIGHTNESS}
\label{sec:result}

We analyze the supercluster sample for each model at each redshift to determine the mean specific sizes of 
the supercluster spines, by the prescriptions described in \citet{SL13} and \citet{shim-etal14}: 
\begin{itemize}
\item[{\em i })] Apply the minimum spanning tree (MST) algorithm \citep{barrow-etal85,colberg07} to the supercluster 
sample from the simulations of \citet{baldi-etal14} to determine a MST of each supercluster. 
\item[{\em ii })] From each supercluster MST, prune away repeatedly the minor twigs having less than three nodes until 
the main stem of each supercluster (i.e. the ``supercluster spine") is determined and count the number of halos belonging to 
the main stem (i.e. the ``supercluster nodes"). 
\item[{\em iii })] Select only those superclusters whose spines consist of three or more nodes. 
\item[{\em iv })] Find a cuboid which fits the spatial distribution of the nodes of each supercluster spine and 
measure the length of its diagonal line as the size of each supercluster spine \citep[see][]{PL09}. 
\item[{\em v })] Determine the specific size ($\tilde{S}$) of each supercluster spine by dividing the size by the node 
number ($N_{\rm node}$). The supercluster spines whose shapes are closer to straight filaments are expected 
to have larger specific sizes.
\item[{\em vi} )] Calculate the mean value of $\tilde{S}$ averaged over all supercluster spines.
\end{itemize}

\begin{figure}[ht]
\begin{center}
\plotone{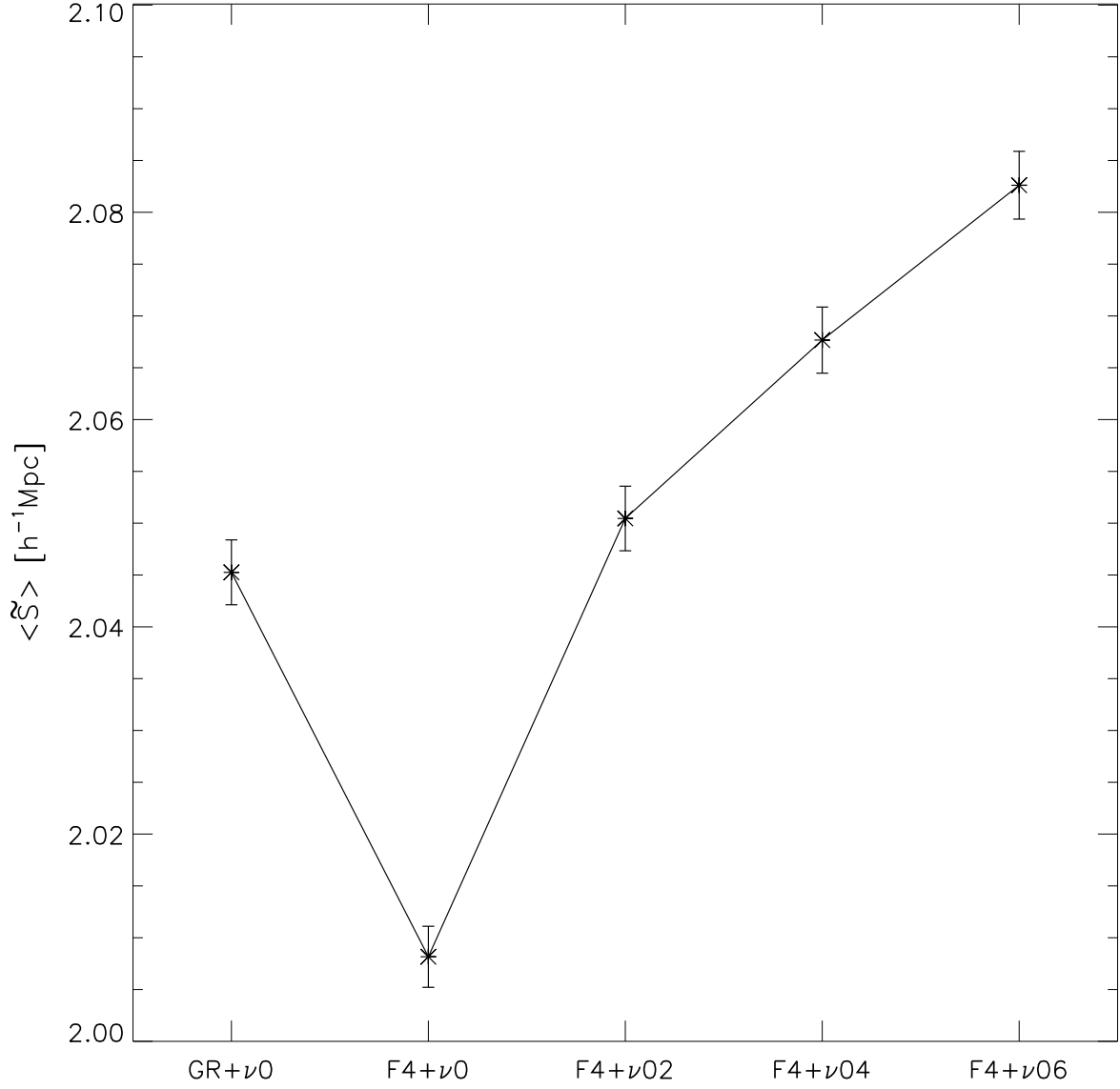}
\caption{Mean specific sizes of the supercluster spines for five different models at $z=0$. The errors represent 
one standard deviation in the measurement of the mean value. }
\label{fig:ssize}
\end{center}
\end{figure}

Figure \ref{fig:ssize} shows the mean specific sizes $\langle\tilde{S}\rangle$ of the supercluster spines for the 
different cosmological models under investigation. The errors are calculated as one standard deviation in the 
measurement of the mean value. The value of  $\langle\tilde{S}\rangle$ is shown to monotonically increase as the 
value of $m_{\nu}$ increases from $0$ to $0.6\,$eV. Interestingly, it is not the F4+$\nu{04}$ model but the 
F4+$\nu{02}$ model which predicts almost the same $\langle\tilde{S}\rangle$ as the $\Lambda$CDM cosmology. In 
fact, between the $\Lambda$CDM and the F4+$\nu{04}$ we find a statistically significant difference in the value of 
$\langle\tilde{S}\rangle$. Given that for the standard LSS statistics \citet{baldi-etal14} found that the F4+$\nu{04}$ 
model appears to be hardly distinguishable from the GR+$\Lambda$CDM standard scenario, our results indicate that 
the value of $\Sigma _{i}m_{\nu _{i}}$ required to compensate for the effect of the fifth force on the specific sizes of the 
supercluster spines and on the basic statistics of LSS are different  from each other. This result leads us to expect that 
it may be possible to break the cosmic degeneracy between the GR+$\Lambda$CDM and the F4+$\nu$ models if the 
basic statistics of LSS are combined with the specific size distribution of the supercluster spines.

\begin{figure}[ht]
\begin{center}
\plotone{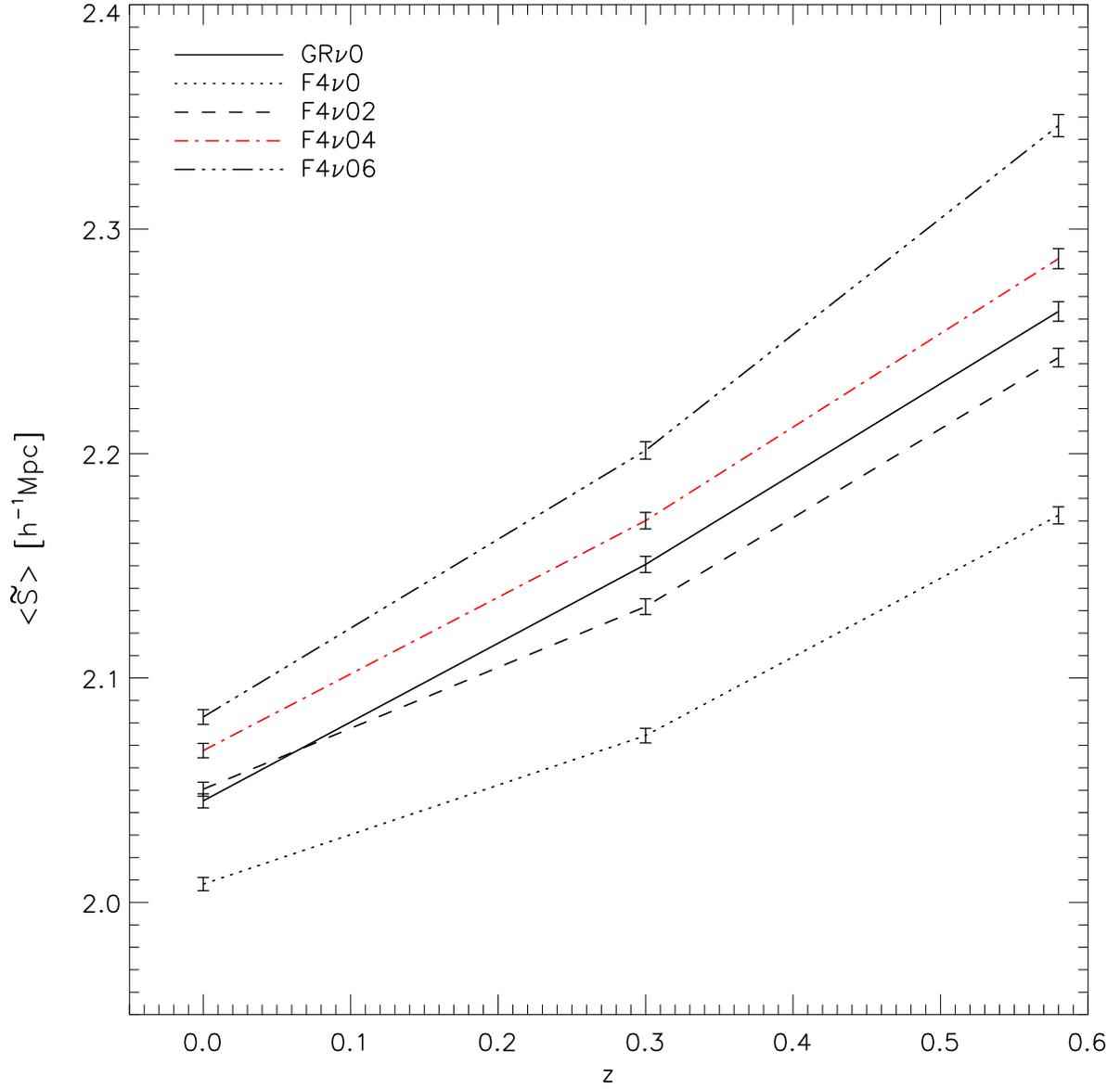}
\caption{Evolution of the mean specific sizes of the supercluster spines for five different models.}
\label{fig:ssevol}
\end{center}
\end{figure}

Figure \ref{fig:ssevol} displays the mean specific sizes of the supercluster spines versus redshifts for the five models.
As for the previous figures, the (red) dot-dashed line corresponds to the {\it most degenerate} combination of the models in the 
standard LSS statistics \citep[as found by ][]{baldi-etal14}.  
As the plot shows, the difference in $\langle\tilde{S}\rangle$ among the fives models becomes larger at higher 
redshifts, which is consistent with the result of \citet{shim-etal14} who showed that  the effect of $f(R)$ gravity on the 
specific size distribution becomes larger at higher redshifts, and this trend appears in all the models under investigation. 
The F4+$\nu{06}$ model exhibits the most rapid evolution of $\langle\tilde{S}\rangle$, which implies that for more massive 
neutrinos the free streaming has a stronger effect of sharpening the filamentary structures when they were more energetic at higher 
redshifts.  In other words, although it is impossible to distinguish between the GR+$\Lambda$CDM and the F4+$\nu{02}$ 
models by measuring $\langle\tilde{S}\rangle$ at the present epoch, this degeneracy is broken by looking at higher redshifts. More 
specifically, it appears possible in principle to distinguish the two models by comparing the value of $\langle\tilde{S}\rangle$ at 
$z\ge 0.3$ when the F4+$\nu{02}$ model significantly deviates from the GR+$\Lambda$CDM standard case. 

\section{DISCUSSION AND CONCLUSION}\label{sec:con}

This work has been inspired by two recent publications. 
On one hand, the work of \citet{shim-etal14} demonstrated that the specific size of the supercluster spine is a powerful 
diagnostic not only for detecting any $f(R)$ modification of gravity but also for distinguishing the effects of $f(R)$ 
gravity from that of the presence of interacting Dark Energy. On the other hand, the recent work of \citet{baldi-etal14} raised for the first 
time a cosmic degeneracy problem consisting in the fact that the basic statistics of LSS (such as e.g. the
cluster mass function, the two-point correlation function of the density field and the halo bias) might inherently fail to 
distinguish a suitable combination of a $f(R)$ MG model and of a specific value of the total mass of neutrinos from the 
standard $\Lambda$CDM cosmology, since the free streaming effect of massive neutrinos effectively suppresses 
the extra clustering effect associated with the scalar fifth-force of $f(R)$ gravity. 

Using the numerical data from the N-body simulations carried out by \citet{baldi-etal14} and the techniques developed 
by \citet{shim-etal14}, we have determined the mean specific sizes of the supercluster spines at three different 
redshifts ($z=0,\ 0.3,\ 0.6$)  for an $f(R)$ gravity cosmology with different values of the total neutrino mass. 
Our investigation showed that the neutrino mass required to maximally suppress the effects of the MG fifth force on the specific size 
distribution of the supercluster spines at $z=0$ is different from the value that determines the maximum degeneracy 
for the basic statistics of LSS. Furthermore, we have also found that at higher redshifts all the considered combined models show 
larger differences in the mean specific size of the supercluster spine from the standard GR+$\Lambda$CDM cosmology. 
Therefore, we conclude that the evolution of the specific size of the supercluster spine can in principle play a crucial role in 
breaking the cosmic degeneracy highlighted by \citet{baldi-etal14}.

Although such conclusion emerges clearly from the present study, a more thorough investigation of the power of the 
supercluster spines in breaking the MG-massive neutrinos degeneracy will be required in order to quantify precisely 
the gain of discriminating power obtained by combining this new statistics with the standard LSS probes.
First of all, the datasets employed for our analysis were obtained from a suite of intermediate-resolution $N$-body 
simulations. Since the identification of the superclusters -- especially at high redshifts -- is likely to be affected by the 
resolution of the original $N$-body realization, a new set of halo catalogs from higher resolution simulations is required 
to examine more carefully how the mean specific sizes of the supercluster spines evolve with redshifts. Secondly, in 
the current work we have considered  only the case of a rather ``extreme" (although still possibly consistent with standard LSS 
data for a reasonable value of the neutrino mass) $f(R)$ scenario, namely $f_{R0}=-10^{-4}$. Therefore, it will be 
necessary to examine whether the specific size of the supercluster spine is capable of breaking the cosmic degeneracy even 
for the more challenging case of less severe deviations from the standard GR gravity.

Second of all, our conclusion is based not on a realistic observational error but only on the statistical significance of the differences in 
the mean specific size of the superclusters spine among the models.  It will be essential to estimate  an observational confidence 
region around the fiducial model for the specific size of the supercluster, as \citet{baldi-etal14} did for the LSS statistics.  
Another direction in which some improvements are needed is the development of an analytic model for the mean 
specific size of the supercluster spine. One of the reasons why the basic LSS statistics have been so widely employed as 
a test for cosmology is that they can be analytically (or semi-analytically) predicted. On the contrary, the mean specific sizes of 
the supercluster spines have so far been only numerically determined, without being guided by any analytic 
prescription, which is an obvious drawback that has to be overcome for its practical application in the future.
We leave these three main extensions of our analysis for future work.

\acknowledgments

JS and JL were financially supported by the Basic Science Research Program through the National 
Research Foundation of Korea(NRF) funded by the Ministry of Education (NO. 2013004372) and 
by the research grant from the National Research Foundation of Korea to the Center for 
Galaxy Evolution Research  (NO. 2010-0027910). 
MB is supported by the  Marie Curie Intra European Fellowship
``SIDUN"  within the 7th Framework  Programm of the European Commission
and also acknowledges financial contributions from the PRIN INAF 2012 ``The Universe
in the box: multi-scale simulations of cosmic structure".
The numerical simulations presented in this work have been performed 
and analyzed on the MareNostrum3 cluster at the BSC supercomputing centre in Barcelona 
(through the PRACE Tier-0 grant ``SIBEL1") and on the Hydra cluster at the RZG supercomputing centre in Garching.

\clearpage
\clearpage
\clearpage

\end{document}